\documentclass{actapoly}
\pdfoutput=1

\usepackage{xcolor}

\DeclareMathOperator{\imag}{Im}
\DeclareMathOperator{\real}{Re}

\newcommand{\ii}{\ensuremath{\mathrm{i}}}

\begin{document}

\title[BEC with $\cal {PT}$-Symmetric Double-Delta Loss and Gain: Test of 
Estimates]
{A Bose-Einstein Condensate with $\cal {PT}$-Symmetric Double-Delta Function Loss and Gain in a Harmonic Trap: A Test of Rigorous Estimates}

\institution{my}{Institut f\"ur Theoretische Physik 1, Universit\"at Stuttgart,
  70550 Stuttgart, Germany}

\author[D. Haag]{Daniel Haag}{my}
\author[H. Cartarius]{Holger Cartarius}{my}
\correspondingauthor[G. Wunner]{G\"unter Wunner}{my}
{wunner@itp1.uni-stuttgart.de}

\begin{abstract}
We consider the linear and nonlinear Schr\"odinger equation for a Bose-Einstein
condensate in a harmonic trap with $\cal {PT}$-symmetric double-delta function 
loss and gain terms. We verify that the conditions for the
applicability of a recent proposition by Mityagin and Siegl on
singular perturbations of harmonic oscillator type self-adjoint operators
are fulfilled. In both the linear and nonlinear case we calculate
numerically the shifts of the unperturbed levels with quantum numbers
$n$ of up to 89 in dependence on the 
strength of the non-Hermiticity and compare with rigorous estimates derived
by those authors. We confirm that the predicted $1/n^{1/2}$ estimate
provides a valid upper bound on the the shrink rate of the numerical 
eigenvalues. Moreover, we find that a more recent estimate of $\log(n)/n^{3/2}$
is in excellent agreement with the 
numerical results. With nonlinearity the shrink rates
are found to be smaller than without nonlinearity, and the rigorous estimates,
derived only for the linear case, are no longer applicable.

\end{abstract}


\maketitle

\section{Introduction}

Bose-Einstein condensates with $\mathcal {PT}$-symmetric loss and gain
have been proposed~\cite{Klaiman08a} as a first experimental realisation 
of $\mathcal {PT}$ symmetry in a real quantum system. 
The idea is to confine the condensate in a double well potential,
and to create a  $\mathcal {PT}$-symmetric situation by coherently injecting
atoms into one well and removing them from the other.
A particular difficulty in the theoretical treatment is that, on account 
of the s-wave scattering of the atoms, the
Gross-Pitaevskii equation, which describes the condensates, 
contains a term $g|\psi|^2$, and thus is nonlinear in the wanted wave function. In a series of papers 
both for realistic set-ups  as well as  for delta-function
models of the double wells
we have shown \cite{Cartarius12b,Cartarius12c,Cartarius13a,Dast13a,Dast13b,Kreibich13a}
 that the nonlinearity introduces new features in 
the evolution of the eigenvalue spectrum as the non-Hermiticity
is increased but yet $\mathcal {PT}$ symmetry of the wave function
is preserved if both nonlinearity and non-Hermiticity are not too
strong.

Bose-Einstein condensates are usually trapped in harmonic
potentials produced by counterpropagating laser beams. 
Therefore condensates with an additional
$\mathcal {PT}$-symmetric double well potential can be regarded
as a ''perturbation'' of the  harmonic oscillator. Aducci and Mityagin 
\cite{Adduci2012} and Siegl and Mityagin \cite{Mityagin2013} have
recently analysed perturbations of harmonic-like operators from
a mathematical point of view. The second paper in particular
also allows for singular perturbations, such as delta-functions.

These authors have proved that the eigenvalues of the perturbed operator
become eventually simple and the root system forms a Riesz basis.
Their results are valid if the following criterion is fulfilled:
\begin{equation}
\forall \, m, n \in 
\mathbb{N} \quad |\langle \psi_m |\hat B|\psi_n\rangle| \le \frac{M}{m^{\alpha} n^{\alpha}},\; \alpha > 0,  M>0 \,,
\label{eq:ms_criterion}
\end{equation}
where $\hat B$ is the perturbation operator, $M$ is a constant
that depends on the type of the perturbation,
 and the $\psi_m$ are the
harmonic oscillator eigenfunctions.
A further prediction is that the shrink rate of the disks into which
the eigenvalues can be shifted is proportional to $1/n^{2\alpha}$, with 
$\alpha$ the exponent appearing in \eqref{eq:ms_criterion}.

It is the purpose of this paper to test the estimates numerically for
the example of a Bose-Einstein condensate in a double well
potential confined by a harmonic trap. To this end we consider
the model of two $\mathcal {PT}$-symmetric delta-function wells,
since in this case simple analytical estimates can be obtained, whereas 
in the case of the realistic double well discussed in  Refs. \cite{Cartarius13a,
Dast13a} complicated estimates in terms of hypergeometric 
functions result. 

It must be mentioned that of course because of their simplicity 
delta functions  have widely been used in the literature in the 
context of $\mathcal {PT}$ symmetry. Spectral properties 
of scattering and bound states in $\mathcal {PT}$-symmetric double- and multiple-delta function potentials have been
investigated e.g. in Refs \cite{Jones1999,Ahmed2001,Demiralp2005a,Demiralp2005b,Mostafazadeh2006a,Mostafazadeh2009a,Mehri-Dehnavi2010a,Uncu2008,Cartarius12b,Cartarius12c}. In all these
papers no external potential was present, in additon to  the delta potentials.

A paper in which $\mathcal {PT}$-symmetric point interactions
were studied embedded in an external potential is that by Jakubsk{\'{y}} and Znojil
\cite{Jakubsky05} who positioned the delta functions in an infinitely
high square well and analysed the spectrum in dependence on the
position of the delta functions within the well. Their work was extended
by Krej{\v{c}}i{\v{r}}{\'{i}}k and Siegl \cite{Krejcirik2010,Siegl2011} who replaced the delta functions by 
$\mathcal {PT}$-symmetric Robin boundary conditions at the edges of the square well. We note that the spectrum of a harmonic oscillator perturbed
by two identical real-valued point interactions has been analysed
by Fassari and Rinaldi \cite{Fassari2012}. 
But to our knowledge the situation of two $\mathcal {PT}$-symmetric
delta functions
in an external harmonic potential has not yet been investigated. 
While in the square well the delta functions can be placed only
within the well, the harmonic oscillator potential has the advantage 
that the delta functions in principle can be shifted to any position 
on the real axis. 

\section{Bose-Einstein condensate in a $\mathcal {PT}$-symmetric
harmonic trap}

At low temperatures and densities Bose-Einstein condensates are well 
described by the Gross-Pitaevskii equation \cite{Gross61a,Pitaevskii61a}
\begin{equation}\label{eq:GPE}
\left( -\frac{d^2}{dx^2} + V(x) + g \vert \psi \vert^2 \right)\psi=\mu\psi.
\end{equation}
Here $\psi$ denotes the condensate wave function, the eigenvalue $\mu$ is 
the chemical potential, and $V(x)$ is the trapping potential to confine
the condensate. The nonlinear
term in (\ref{eq:GPE}) arises from the s-wave scattering
interaction of the atoms; $g$ is a measure for the strength of this 
interaction. We consider a harmonic trapping potential 
and model a $\cal {PT}$-symmetric
double well with equilibrated loss and gain by imaginary delta functions. 
Thus the Hamiltonian we consider is given by
\begin{equation}\label{eq:Ham}
{\hat H} =  -\frac{d^2}{dx^2} + x^2 + \ii\gamma\Big(\delta(x-b)-\delta(x+b)\Big)
+g |\psi(x)|^2 \,.
\end{equation}
In (\ref{eq:Ham}) $\pm b$ denotes the position of the imaginary deltas
and $\gamma$ the strength of the non-Hermiticity 
We will later consider the effects of the nonlinearity 
on the spectrum, but for the time being 
assume that the nonlinearity is negligible in order to be in a position 
to compare with the predictions of Mityagin and Siegl \cite{Mityagin2013}.

\begin{figure}
\centering
\includegraphics[width=1.0\linewidth]{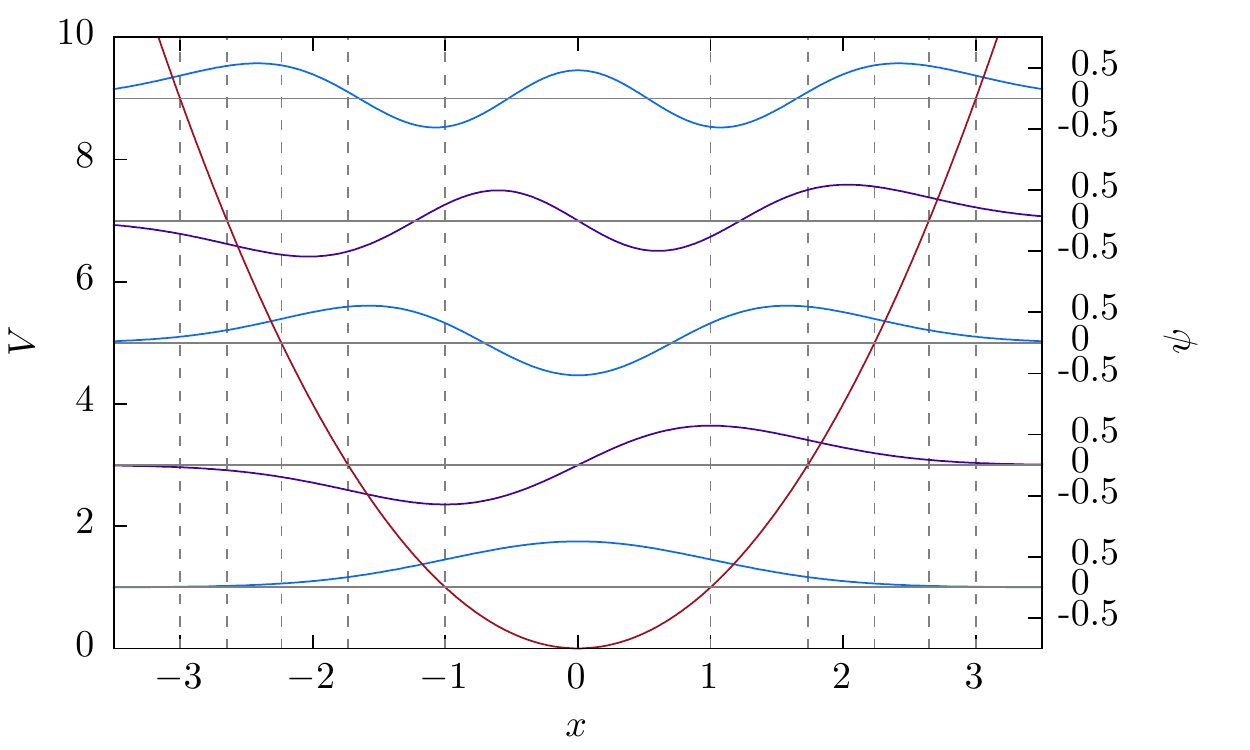}
\caption{Unperturbed spectrum of the harmonic oscillator and the lowest
five wave functions. Vertical dashed lines designate different positions
of the $\mathcal {PT}$-symmetric delta-function perturbations.}
\label{figure1}
\end{figure}

The eigenvalues of the unperturbed spectrum are given by $\mu_n = 2n+1$ ($n = 
0, 1, 2, \dots) $. 
Figure \ref{figure1} shows the unperturbed spectrum together
with the wave functions of the lowest five states. The dashed vertical
lines designate different positions at which the delta functions are
placed. From the figure it is already obvious that only such states
will be significantly affected by the perturbation
which are within the classically allowed region at the positions the 
delta functions.
By contrast, states for which the delta functions lie in the  
classically forbidden (exponentially decaying) regime 
will not be affected. This means that as the delta functions 
are shifted further and further out an increasing number of low-lying
eigenvalues will not be changed by the perturbation. 

It is easy to show that the Mityagin-Siegl criterion
\eqref{eq:ms_criterion} is fulfilled for the imaginary delta function
perturbation in \eqref{eq:Ham}
since
\begin{equation}\label{eq:perturbation}
|\langle \psi_m |\hat B|\psi_n\rangle| = \\ |\gamma|\,|\psi_m(b)\psi_n(b) - \psi_m(-b)\psi_n(-b)|.
\end{equation}
The oscillator eigenfunctions have either even or odd parity,
therefore $|\langle \psi_m |\hat B|\psi_n\rangle|= 0$ if
$m$ and $n$ are both even or odd, while for $m$ even and $n$ odd,
and vice versa, we have 
\begin{eqnarray}\label{eq:estimate}
|\langle \psi_m |\hat B|\psi_n\rangle|&=&  
2|\gamma| |\psi_m(b)| |\psi_n(b)|\\
&\le& 2|\gamma|\; {\tilde C}\, m^{-1/4}n^{-1/4}.
\end{eqnarray}
In the last line we have exploited an inequality
given by Mityagin and Siegl \cite{Mityagin2013} which is valid for 
$2(2n+1)  \ge b^2$. The prediction
then is that eigenvalues can only move 
to a distance $M$, where $M$ is a constant, uniform for all eigenvalues,
and 
in particular, that there exists an $n_0$ such that for $n \ge n_0$ all 
eigenvalues stay in disjoint disks with shrinking radii, with the shrink rate
being bounded from above by $C n^{-1/2}$. In fact, one of the authors of Ref. \cite{Mityagin2013} 
has pointed out \cite{Mityagin2014} that for the case of two 
delta potentials the shrink rate can be estimated even more precisely
to behave as $\log(n)/n^{3/2}$.
Note that no such statements can be made from their theorems 
for the case that the nonlinearity
is also included as a perturbation.

\section{Eigenvalue spectra}

The (real and complex) eigenvalues $\mu$ of the Hamiltonian \eqref{eq:Ham}
and its eigenstates $\psi$ are obtained, both in the linear and the nonlinear
case, by integrating the wave functions outward from $x=0$, and varying
the initial values of $\real \psi(0)$, $\psi^\prime(0) \in \mathbb{C}$, and
 $\mu \in \mathbb{C}$ to find  square-integrable normalised solutions
(the arbitrary global phase is exploited by 
choosing $\imag \psi(0) = 0$).

\begin{figure}
\centering
\includegraphics[width=1.0\linewidth]{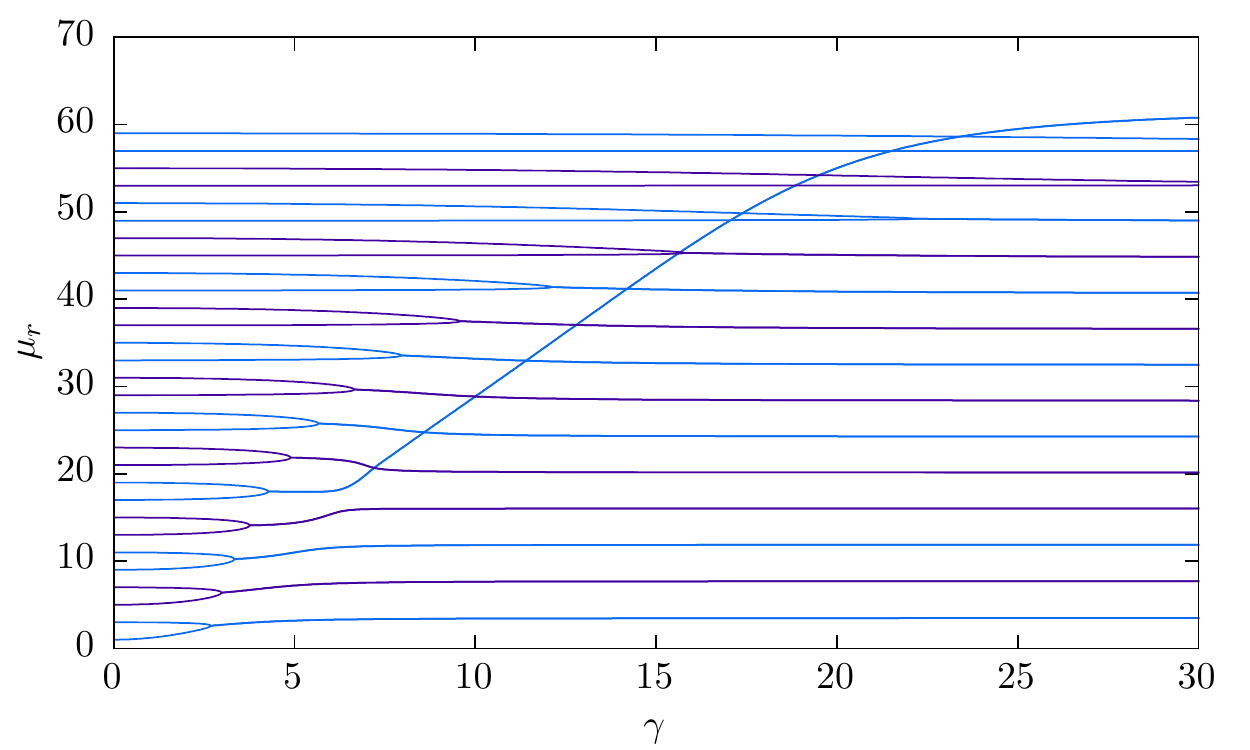}
\\[1mm]
\includegraphics[width=1.0\linewidth]{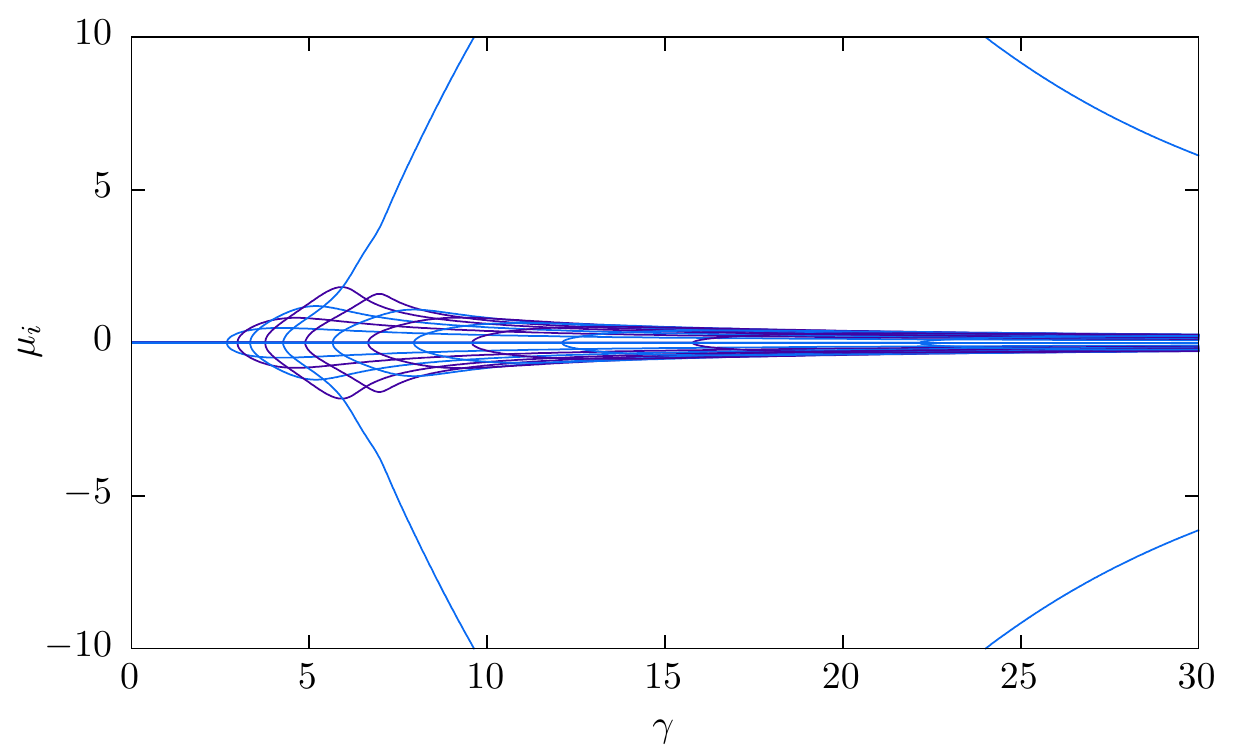}
\caption{Real parts (top) and imaginary parts (bottom) of the eigenvalues
of \eqref{eq:Ham} (with $g = 0$) for $b=0.2$, evolving 
from the unperturbed levels with $n= 0 , \dots ,29$, in dependence of the 
strength $\gamma$ of the non-Hermiticity.}
\label{figure2}
\end{figure}

Figure \ref{figure2}  
shows  for the lowest 30 levels the real and imaginary
parts of the eigenvalues of \eqref{eq:Ham} (for $g = 0$) as functions
of the strength $\gamma$ of the non-Hermiticity, for a position of the
delta functions at $b = 0.2$, close to the centre of the oscillator.
One recognises that successive pairs of eigenvalues coalesce at
branch points, from whereon they turn into complex conjugate pairs.
The branch points are shifted to larger values of $\gamma$ as one goes
up in the spectrum. What is surprising is that both the real and imaginary
part of the complex eigenvalues emerging from the branch point
of the eighth and ninth excited state experience huge shifts, but eventually
saturate, like the other levels. For the latter, the real parts remain approximately
constant beyond the branch points, and the imaginary parts 
quickly tend to zero.

\begin{figure}
\centering
\includegraphics[width=1.0\linewidth]{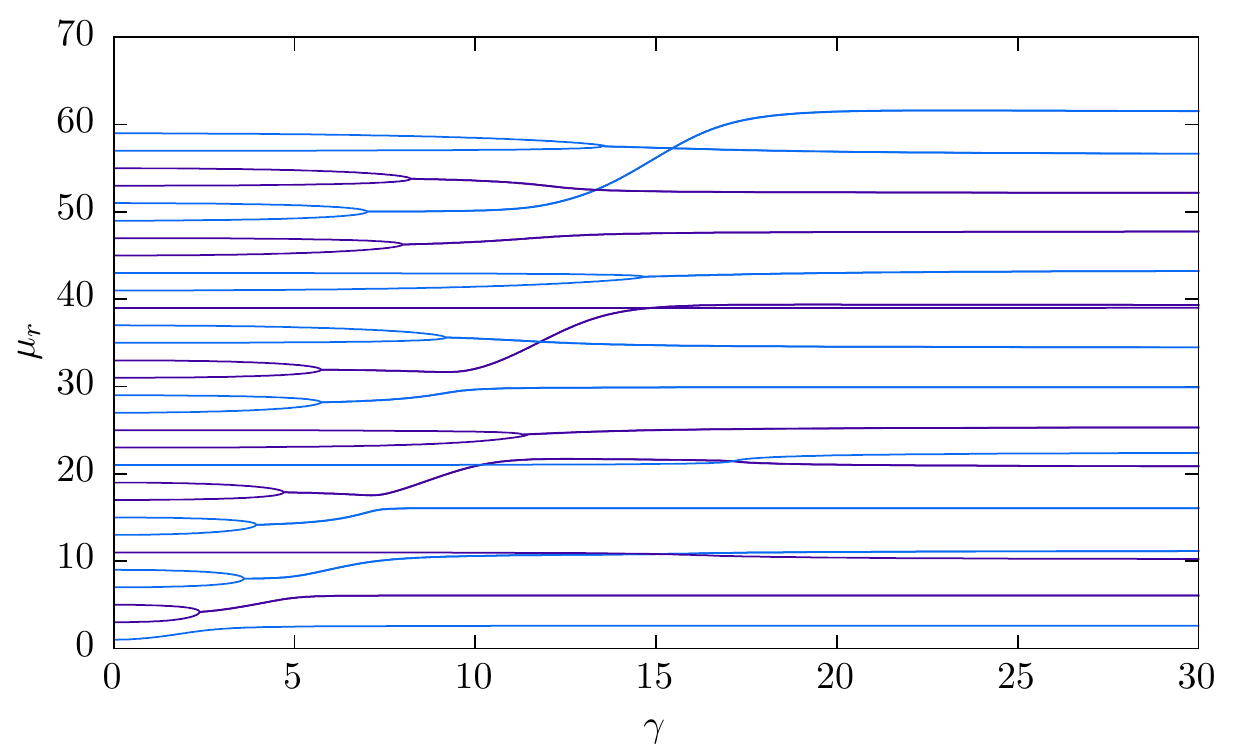}
\\[1mm]
\includegraphics[width=1.0\linewidth]{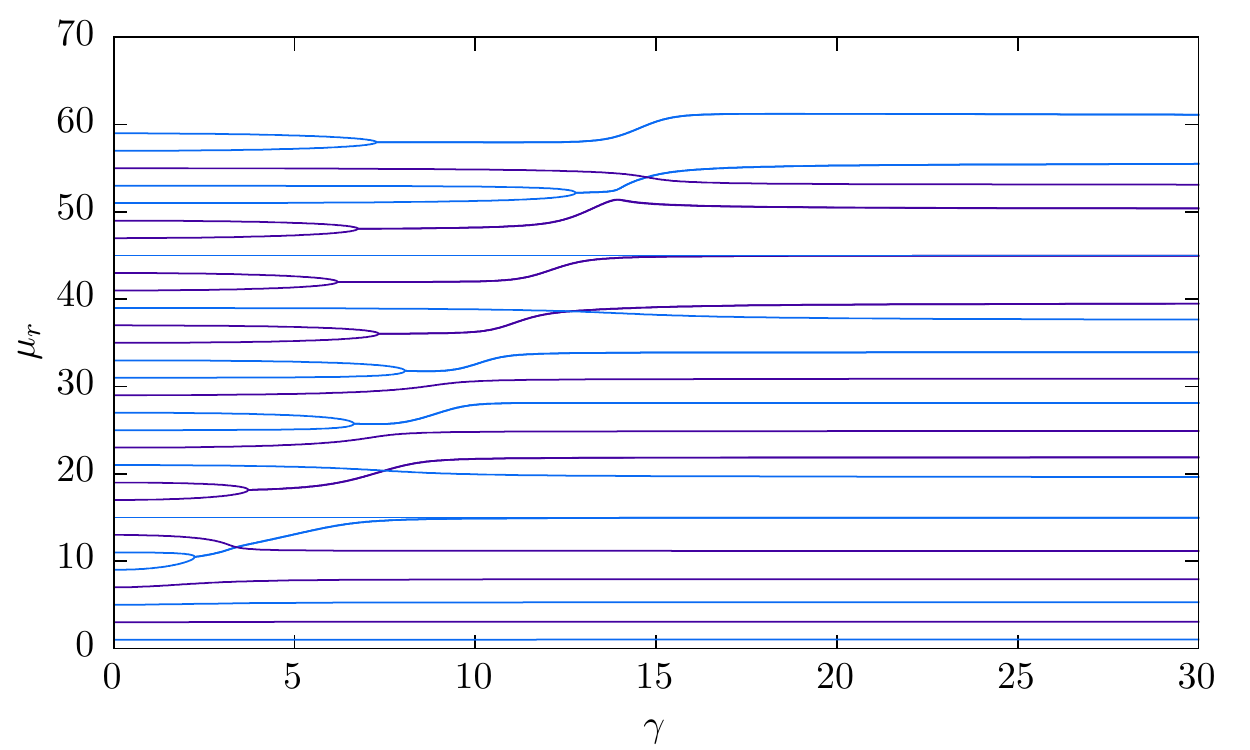}
\caption{Real parts  of the eigenvalues
of \eqref{eq:Ham} (with $g = 0$), evolving 
from the unperturbed levels with $n= 0 , \dots ,29$, 
with the delta functions placed into the classical turning point of the
unperturbed ground state ($b = 1$, top) and the third excited state 
($b=\sqrt{7}$, bottom).
}
\label{figure3}
\end{figure}

In Figure \ref{figure3}  
we move the delta functions further away from the centre of 
the harmonic oscillator to the classical turning points 
$b = 1$ and $\sqrt{7}$ of the unperturbed levels with $n =0$ and  $n=3$,
respectively. At the turning points the unperturbed 
wave functions enter into the classically forbidden, i.e. exponentially
decreasing region. It is therefore no surprise that for $b =1$ the
ground state no longer ''feels'' the delta functions, and no longer
unites with the first excited state at a branch point. Rather its
eigenvalue remains real for any strength of the non-Hermiticity. 
For  $b=\sqrt{7}$ it is the states with $n=0, 1, 2, 3$ which exhibit this property.

We note that similar behaviour was found  by Jakubsk{\'{y}} and Znojil
 \cite{Jakubsky05}   
in their $\mathcal {PT}$-symmetric square well model.
In their terminology, energy levels which coalesce at a branch point
and turn complex are called ''fragile'', while energy levels whose
eigenvalues remain real for any strength of the perturbation are 
called ''robust''. The latter also include states where the delta functions
happen to be at or close to a node of the wave function, and therefore
remain unaffected by the perturbation. Examples of this can be seen in
Figure \ref{figure3}. 

\begin{figure}
\centering
\includegraphics[width=1.0\linewidth]{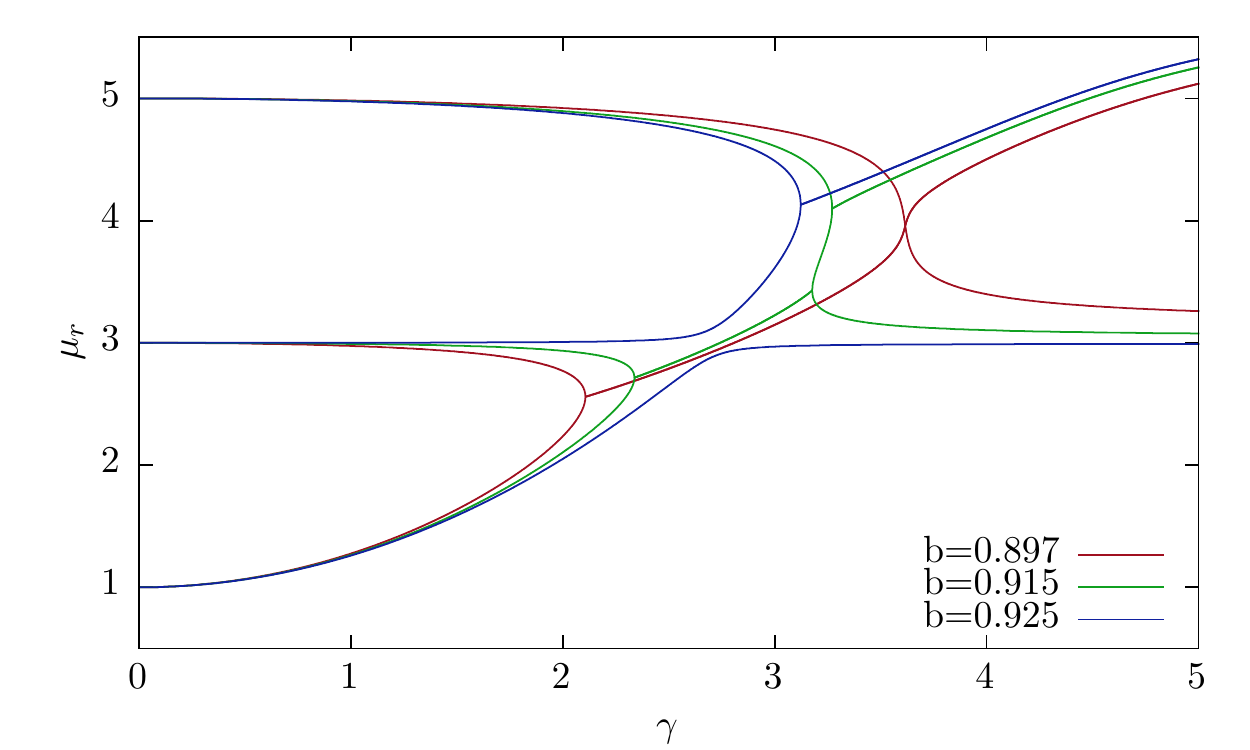}
\includegraphics[width=1.0\linewidth]{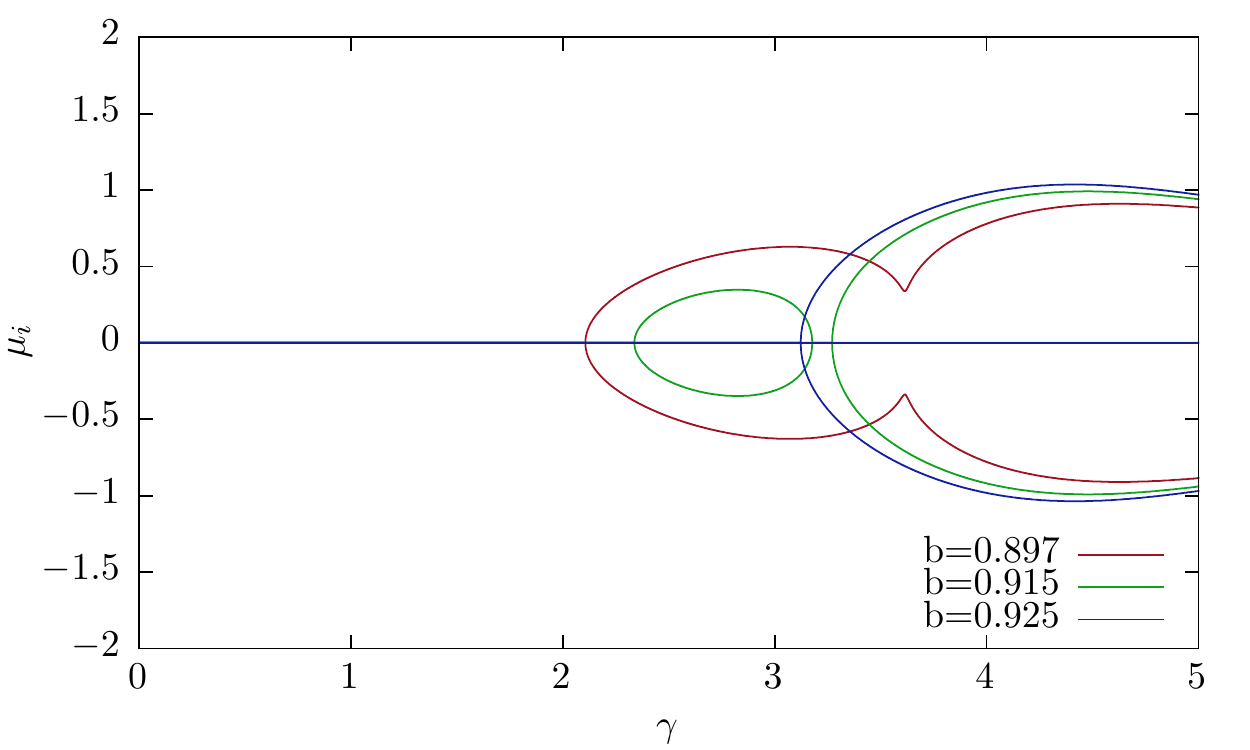}
\caption{Change in the coalescence and bifurcation behaviour of the three 
lowest eigenstates as a function of $\gamma$ at three different positions
of the delta functions  in the vicinity of $b=0.9$. (Top: real parts, bottom: imaginary parts of the eigenvalues.)
}
\label{figure4}
\end{figure}

In Figure \ref{figure2} the ground state coalesces with the first excited
state, while in Figure \ref{figure3}, at $b=1$ (top panel), it has become a single real
level for any value of $\gamma$, and the first excited state coalesces with the 
second excited one. The question arises how the transition
between the different coalescence behaviour  occurs. This is illustrated in 
Figure \ref{figure4} where the real and imaginary parts
of the eigenvalues emerging from the ground state and the 
first two excited levels are shown as functions of $\gamma$, for three positions of the delta functions around $b \approx 0.9$. 
It is evident that at $b=0.897$ the ground and
first excited state still coalesce, giving rise to a pair of complex
conjugate eigenvalues which ''collides'' with the real eigenvalue of the
second excited level. The latter then turns into another pair of 
complex conjugate eigenvalues. The behaviour is similar for $b=0.915$ 
but here the pair of complex eigenvalues resulting from the
merger of the ground and first excited state disappears by splitting 
into two real eigenvalues the lower of which remains real for any $\gamma$,
while the higher after a small interval of $\gamma$ coalesces with the
real eigenvalue of the second excited state and a new pair of complex eigenvalues is born.
Finally, at $b=0.925$ the transition has occured,
the ground state has become a single real level, and the two first excited
states come together. We note again that similar behaviour
was found by Krej{\v{c}}i{\v{r}}{\'{i}}k and Siegl \cite{Krejcirik2010}
in their studies of eigenvalues in a square well with Robin boundary 
conditions (cf. Figure~6 in \cite{Krejcirik2010}).

\begin{figure}
\centering
\includegraphics[width=1.0\linewidth]{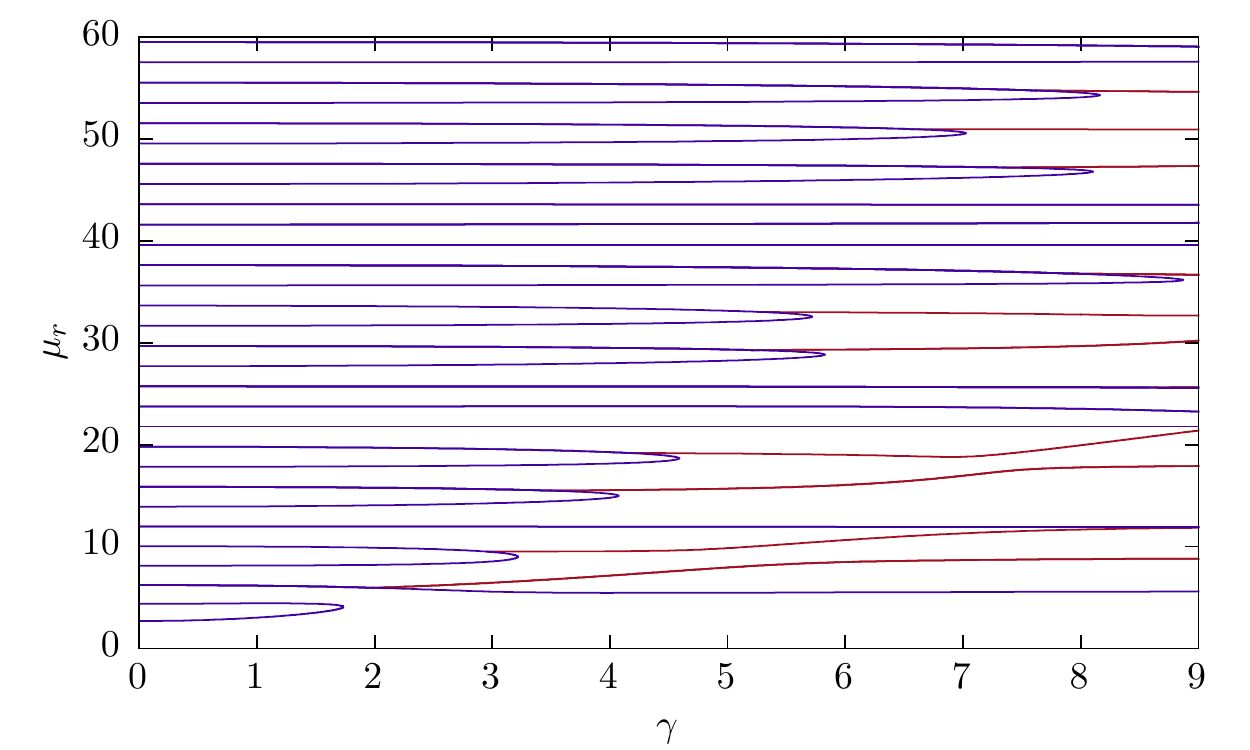}
\caption{Real parts of the eigenvalues
of \eqref{eq:Ham} with finite nonlinearity $g = 5$, evolving 
from the unperturbed levels with $n= 0 , \dots ,29$, for $b = 1$
in dependence on the strength of the non-Hermiticity. Blue lines denote
purely real eigenvalues, whereas real parts of complex eigenvalues (single
lines) are shown in red.}
\label{figure5}
\end{figure}

We now proceed to results for nonvanishing nonlinearity.
Figure \ref{figure5} shows the spectrum for a value of $g = 5$ with the
delta functions placed at $b = 1$.  
The overall behaviour is similar (many states coalesce at branch points),
but there are significant differences. Like in our previous studies of 
$\mathcal {PT}$-symmetric Hamiltonians with nonlinearity 
\cite{Cartarius12b,Cartarius12c,Cartarius13a,Dast13a,Dast13b,Heiss13a}, pairs
of complex conjugate eigenvalues (imaginary parts not shown) 
appear {\em before} the branch points are reached. Again there exist ''robust''
levels whose eigenvalues remain real for any value of $\gamma$.

\section{Eigenvalue shifts}

\begin{figure}
\centering
\includegraphics[width=1.0\linewidth]{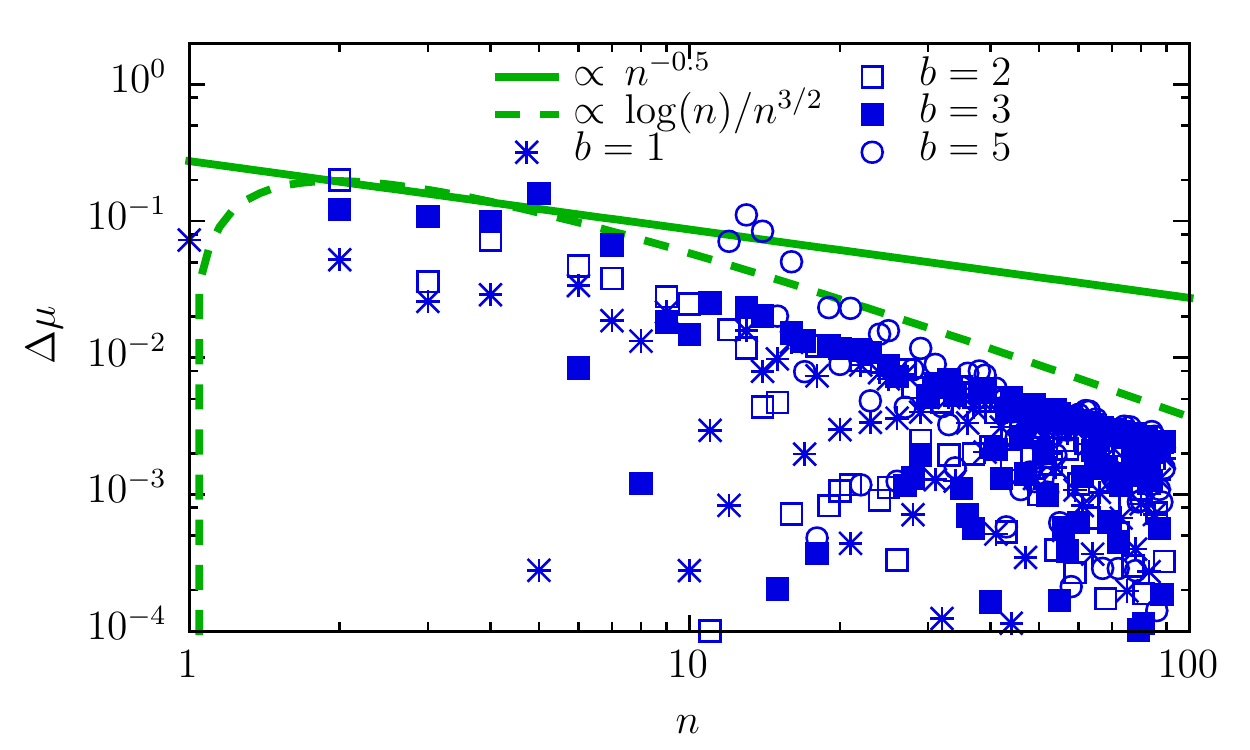}
\\[1mm]
\includegraphics[width=1.0\linewidth]{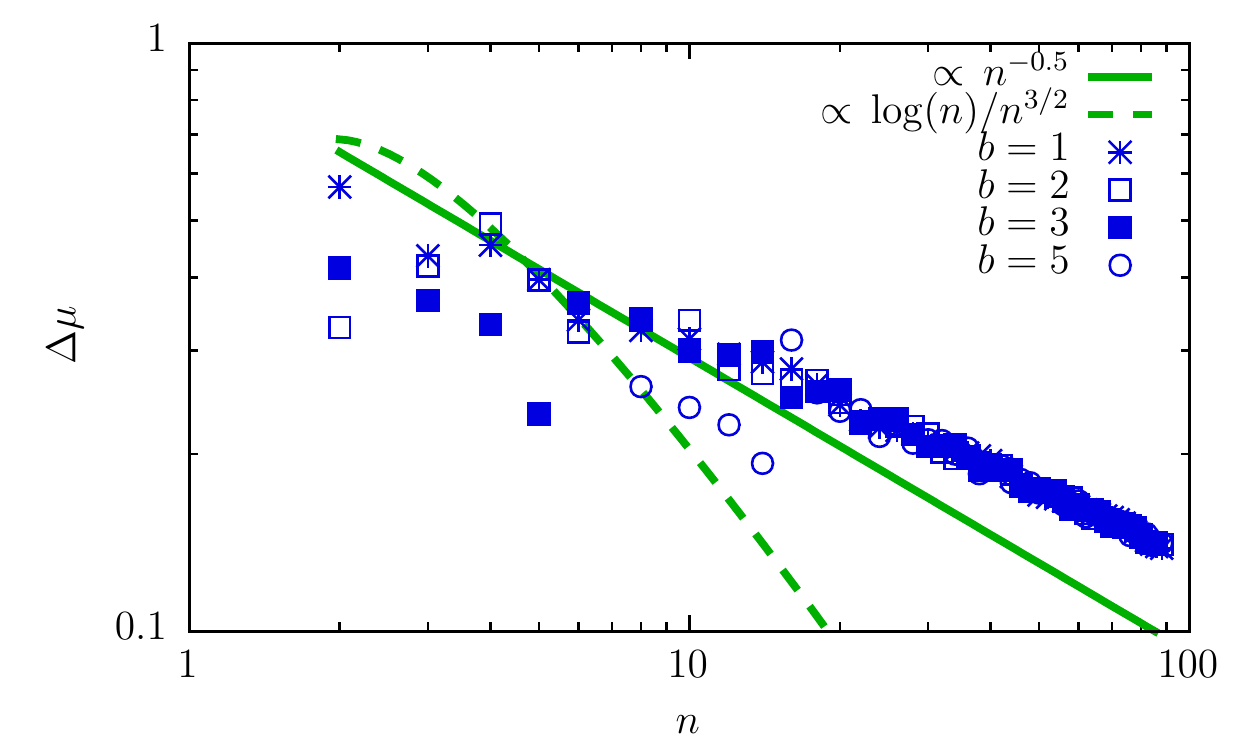}
\caption{Shift of the eigenvalues for
$\gamma =1$ and different positions of the delta functions in comparison
with the mathematically estimated $n^{-1/2}$ and $\log(n)/n^{3/2}$ dependence. 
Top: vanishing nonlinearity, $g = 0$, bottom: $g = 2$.
}
\label{figure6}
\end{figure}

For vanishing nonlinearity, the upper part of Figure \ref{figure6}  
shows, for $\gamma=1$ and different positions of the delta functions,
the shifts of the eigenvalues as a function of the
quantum number $n$
in comparison with the predicted shrink rates of $n^{-1/2}$ of \cite{Mityagin2013} and the improved estimate of $\log(n)/n^{3/2}$ \cite{Mityagin2014}.
It can be recognized that the $n^{-1/2}$ dependence is indeed an upper
bound on the shrink rate, and, moreover, that the improved estimate  is in excellent agreement with the numerical
data, irrespective of the position of the deltas!

The picture changes when the nonlinearity is switched on. 
The bottom part of Figure \ref{figure6} shows eigenvalue shifts
for $g=2$, and different positions 
of the imaginary delta potentials. From the comparison of the scales
of the vertical axes in Figure \ref{figure6} it can be seen that with nonlinearity the
shifts even for the highest states are still bigger than 0.1, while  
they have dropped already well below $10^{-2}$ in the linear case. 
Furthermore, the shrink rate is found to be slower (approximately proportional
to $n^{-0.37}$) than predicted by both rigorous mathematical estimates
for the linear case.  We therefore find significant 
differences in the shrink rates with and without nonlinearity.

For small $\gamma$, all eigenvalues are still real. 
The question therefore suggests itself how the situation changes
when eigenvalues are involved that are shifted into the complex plane,
which happens for increasing $\gamma$. 

\begin{figure}
\centering
\includegraphics[width=1.0\linewidth]{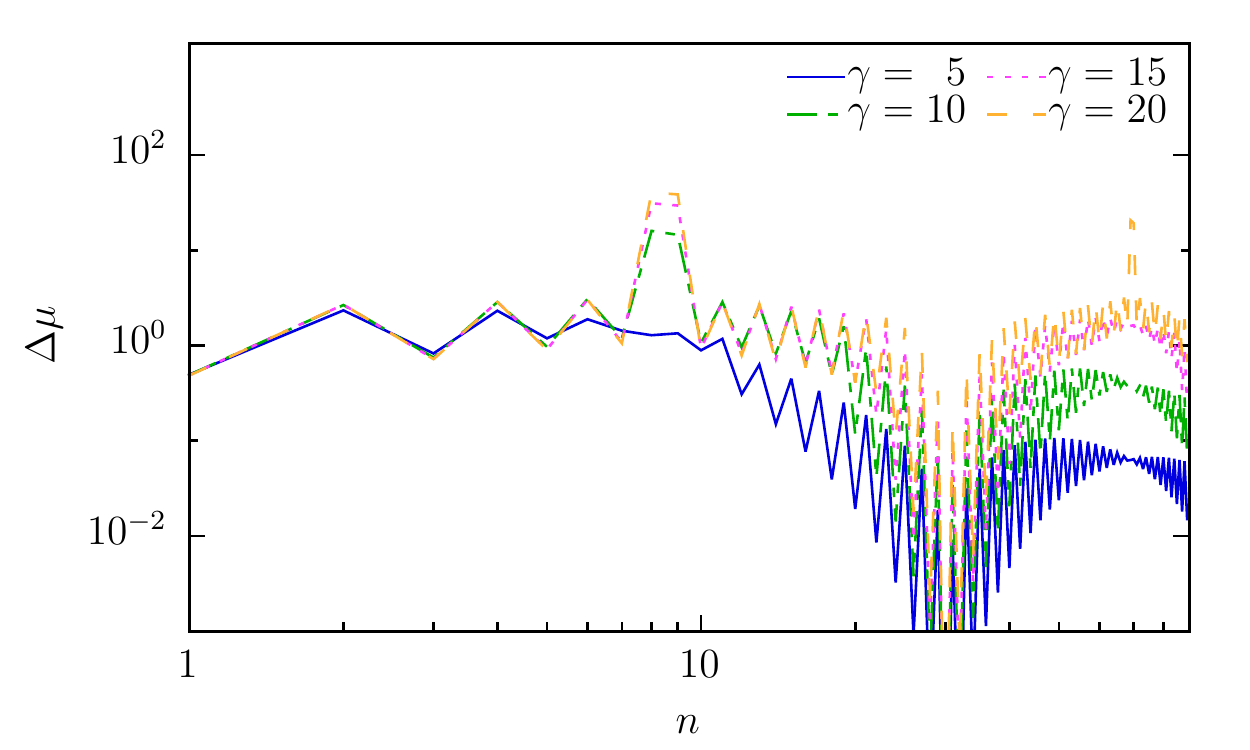}
\caption{Shift of the real and complex eigenvalues 
for growing values of $\gamma$ for $b=0.2$, and $g=0$. 
}
\label{figure7}
\end{figure}

Figure \ref{figure7} 
shows for the eigenvalue spectra with $b = 0.2$, vanishing 
nonlinearity  and different values of $\gamma$ the
distances $|\Delta \mu|$ of the real or complex eigenvalues 
from their original values
(cf. also Figure \ref{figure2}).
The outlier at $n=8$ and $9$ is caused by the giant change of the
real and imaginary parts of the complex eigenvalues beyond the 
branch point of the corresponding states observed already in
Figure \ref{figure2}. A similar outlier occurs around $n=75$,
and a monotonous decrease of the eigenvalue shifts in our calculations
only sets in for $n \approx 80$.

\begin{figure}
\centering
\includegraphics[width=1.0\linewidth]{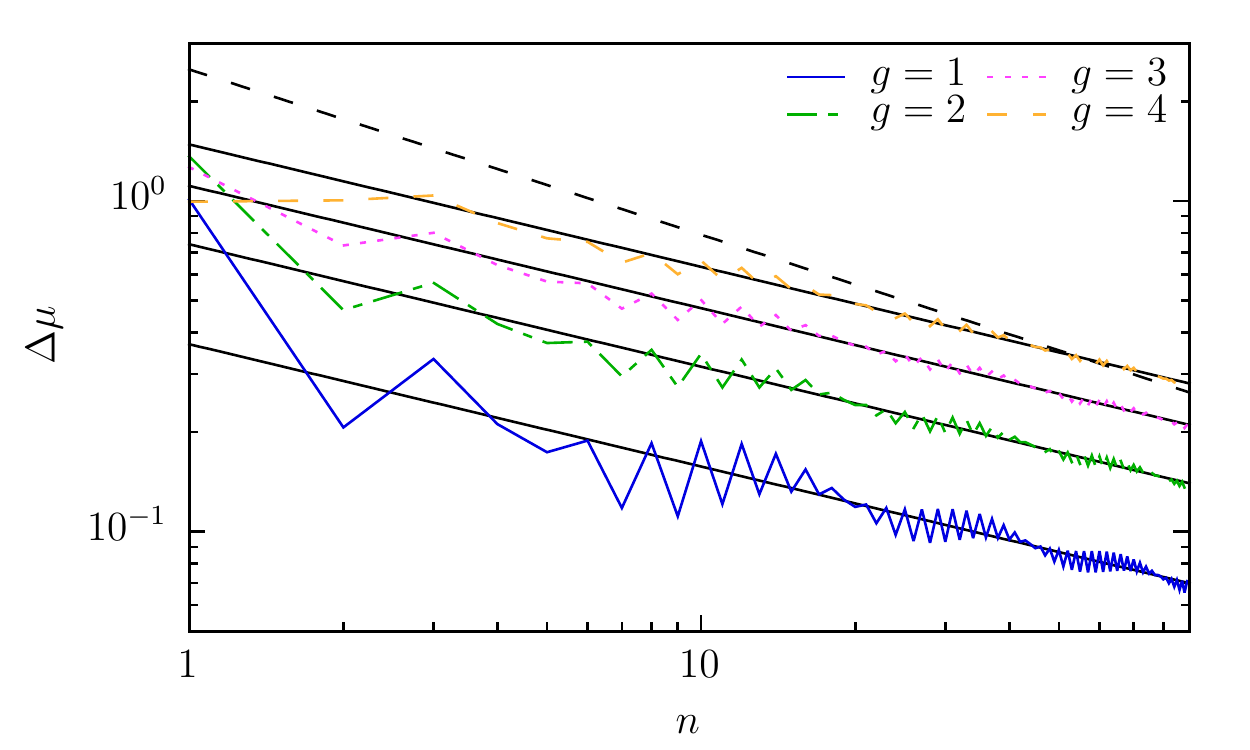}
\\[1mm]
\includegraphics[width=1.0\linewidth]{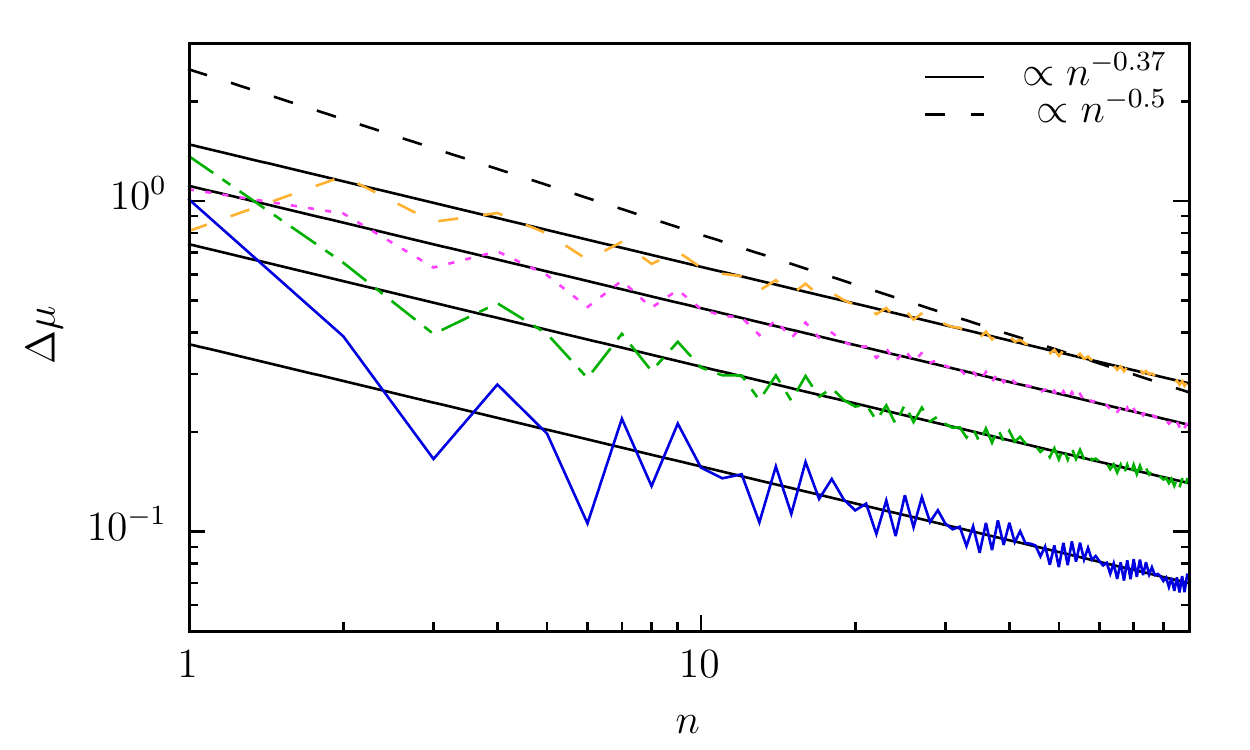}
\caption{Shift of the eigenvalues $N=2n+1$ for four different values
of the nonlinearity at $\gamma=1.5$ for $b=0.5$ (top) and $b=1$
(bottom).}
\label{figure8}
\end{figure}

Finally in Figure \ref{figure8}
we investigate the effect of the nonlinearity on the eigenvalue
shifts for $\gamma=1.5$ and $b=0.5$ and $b=1$. 
We find that the eigenvalue shifts oscillate around
straight lines with slopes -0.37, irrespective of the strength
of the nonlinearity. The amplitude of the oscillations, however, and their
numbers depend on the position of the delta functions. Again the
slope of the estimate proportional to $n^{-1/2}$, also shown in the Figure,
is steeper than the actual slopes found in the numerical results.

\section{Conclusions}
We have carried out a numerical analysis of a
$\mathcal {PT}$-symmetric double delta perturbation of the
harmonic oscillator. We have also considered the case 
were in addition a Gross-Pitaevskii nonlinearity 
proportional to $|\psi|^2$
is present. With the latter, the system can be considered as
a model of a Bose-Einstein condensate in a double well with loss
and gain of atoms. 

We have checked that the Mityagin-Siegl
criterion for the perturbed eigenfunctions to form a Riesz basis
is fulfilled, and compared rigorous mathematical estimates for the
shrink rate of the eigenvalue shifts in dependence on the 
harmonic oscillator quantum number $n$ for various strengths of the
non-Hermiticity and the nonlinearity. We have verified that in 
the linear case the mathematical prediction
for the shrink rates proportional to  $1/n^{1/2}$ is a valid estimate, and 
that the improved estimate proportional to $\log(n)/n^{3/2}$
is in excellent agreement with the behaviour of the 
shrink rates found in the numerical results. 
By contrast, with nonlinearity we find slopes of approximately
$-0.37$, less steep than both  mathematical estimates.
Evidently, to derive estimates also for the nonlinear case
remains a mathematical challenge.

A peculiarity found is the occurrence of outliers in the eigenvalue
spectra which appear beyond branch points
with unusually large real and imaginary parts of their
complex conjugate eigenvalues. They are the reason why for 
growing strength of the non-Hermiticity the asymptotic 
shrink rate behaviour is attained only for high values of  
$n$. Certainly the nature of these outliers and their
mathematical importance should be clarified
in future studies.

\begin{acknowledgements}
We are grateful to Petr Siegl for helpful explanations and discussions.
We are also grateful to Boris Mityagin for very helpful remarks, and
for communicating his improved estimate for the two-delta potentials.
Furthermore we thank two anonymous referees for valuable comments. 
In particular,
one referee points out that a deeper analytic insight 
in the bottom of the spectra could probably be obtained using the
strategies described in Ref. \cite{Fassari2012} for the self-adjoint
case, if adapted to the $\cal{PT}$-symmetric perturbation. 
This certainly also is a useful suggestion for future work.

\end{acknowledgements}

\bibliographystyle{actapoly}

\end{document}